\documentclass[superscriptaddress,amsmath,amssymb,prx,nofootinbib,twocolumn]{revtex4-2}
\usepackage{hyperref}
\usepackage[graphicx]{realboxes}
\usepackage{placeins} 
\usepackage{babel}
\usepackage{times}
\usepackage[usenames,dvipsnames]{color}
\usepackage{soul}
\usepackage[utf8]{inputenc}
\usepackage{float}
\usepackage{dsfont} 

\begin{document}

\newcommand{\REF}[1]{\textcolor{red}{REF(#1)}}
\newcommand{\red}[1]{\textcolor{red}{#1}}
\newcommand{\TODO}[1]{\textcolor{red}{TODO #1}}

\newcommand{\bra}[1]{\left\langle #1 \right\vert }
\newcommand{\ket}[1]{\left\vert #1 \right\rangle }
\newcommand{\ev}[1]{\left\langle #1 \right\rangle }

\newcommand{\sx}{\sigma^x}
\newcommand{\sz}{\sigma^z}

\newcommand{\id}{\mathds{1}}

\title{Constrained dynamics and confinement in the two-dimensional quantum Ising model}

\author{Luka Pave\v{s}i\'{c}}
\email{luka.pavesic@unipd.it}
\affiliation{Dipartimento di Fisica e Astronomia “G. Galilei”, via Marzolo 8, I-35131 Padova, Italy}
\affiliation{Istituto Nazionale di Fisica Nucleare (INFN), Sezione di Padova, I-35131 Padova, Italy}

\author{Daniel Jaschke}
\affiliation{Dipartimento di Fisica e Astronomia “G. Galilei”, via Marzolo 8, I-35131 Padova, Italy}
\affiliation{Istituto Nazionale di Fisica Nucleare (INFN), Sezione di Padova, I-35131 Padova, Italy}
\affiliation{Institute for Complex Quantum Systems, Ulm University, Albert-Einstein-Allee 11, 89069 Ulm, Germany}

\author{Simone Montangero}
\affiliation{Dipartimento di Fisica e Astronomia “G. Galilei”, via Marzolo 8, I-35131 Padova, Italy}
\affiliation{Istituto Nazionale di Fisica Nucleare (INFN), Sezione di Padova, I-35131 Padova, Italy}

\begin{abstract}
We investigate the dynamics of the quantum Ising model on two-dimensional square lattices up to 16 × 16
spins. In the ordered phase, the model is exhibits dynamically constrained dynamics, leading
to confinement of elementary excitations and slow thermalization. After demonstrating the signatures of
confinement, we probe the dynamics of interfaces in the constrained regime through sudden quenches of
product states with domains of opposite magnetization. We find that the nature of excitations can be captured
by perturbation theory throughout the confining regime, and identify the crossover to the deconfining regime.
We systematically explore the effect of the transverse field on the modes propagating along flat interfaces and
investigate the crossover from resonant to diffusive melting of a square of flipped spins embedded in a larger
lattice.
\end{abstract}

\maketitle

While interacting quantum systems typically relax to equilibrium relatively quickly, there are striking examples where this is not the case~\cite{Polkovnikov2011, DeRoeck_2019}. 
Investigating the origins of the anomalously long prethermal regime of such systems has applications in understanding exotic phases of condensed matter~\cite{Yang1989, Moudgalya2020}, phenomena like false vacuum decay~\cite{Coleman1977} and quark confinement~\cite{Greensite2020}.

One possible reason for slow thermalization is the emergence of dynamical constraints which approximately restrict the time evolution to certain subspaces of the Hilbert space.
A prominent example are systems with confinement~\cite{James2019}, where pairs of elementary excitations experience attraction which increases with increasing separation, and thus form long-lived bound states.
The resulting slow thermalization has been extensively studied in Ising spin chains.  
At small transverse fields, the low energy excitations are domain walls, and pairs of them feel a confining force provided by the longitudinal field~\cite{McCoy1978, Kormos2016, Mazza2019, Vovrosh2021, Tan2021, Magoni2021} or long-range interactions~\cite{Liu2019}.

In the two-dimensional (2D) quantum Ising model, the dynamical constraint comes from the lattice itself. 
An excitation has to flip neighbouring spins to propagate, thus creating domain walls. 
At small transverse fields such processes are energetically costly, and thus strongly suppressed.
Confinement has initially been studied in ladders~\cite{Ramos2020, Lagnese2022}, by orthogonally coupling chains described by the Ising field theory in the scaling limit~\cite{Konik2009, James2019} and with infinite tensor network methods~\cite{Tindall2024, Tindall2023}.

In this letter, we use tree tensor networks (TTNs) to explore the constrained dynamics of the quantum Ising model on 2D square lattices of up to $16 \times 16$ spins, through quenches of product states to finite transverse fields.
We demonstrate signatures of confinement for a completely polarized state and investigate the nature of low-energy excitations that drive the dynamics.
Then we turn our attention to inhomogeneous states with domain walls. 
We consider two representative types: a flat interface, and a square domain embedded in a larger lattice.

In the limit of infinite interaction strength, the dynamics of such states is dominated by surface modes which are confined to and propagate along domain walls~\cite{Balducci2022, Balducci2023}. 
We investigate the transition towards diffusive melting and coarsening of interfaces at finite transverse fields, and characterize the nature of domain melting in intermediate regimes.
We conclude by discussing the feasibility of implementing our calculations on quantum simulators and suggest a number of possible extensions of the presented work.

\textit{Model and methods} ---
The Hamiltonian of the two-dimensional quantum Ising model is
\begin{equation}
    H = -J \sum_{\langle i,j \rangle} \sz_i \sz_j - g \sum_i \sx_i,
\end{equation}
with $x$- and $z$-Pauli matrices $\sx$ and $\sz$, $J$ the interaction, and $g$ the transverse field strength. 
Sum indices run over sites of a $N \times N$ square lattice, with $\langle i,j \rangle$ indicating a sum over nearest neighbours.
We work with periodic boundary conditions throughout, but the observed physics does not depend on this choice.

Throughout, we investigate instantaneous quenches of product states with spins aligned in the $z$-direction (eigenstates at $g=0$) to finite values of $g$. 
We encode them into tree tensor networks~\cite{Shi2006, Silvi2019, Cataldi2021} and propagate in real time using the time-dependent variational principle (TDVP)~\cite{Bauernfeind2020, Haegeman2016}, with a timestep of $0.005\ 1/J$.
Our results are generally sufficiently converged for the maximal bond dimension $\chi = 100$. 
See Supplemental Material (SM) for technical details and convergence analysis. 

We characterize the physics through various observables.
Aside from the local magnetization $\ev{\sz_i}$, we use the spin-spin correlation function
$C_{ij} = \langle \sz_{i} \sz_{j}\rangle - \langle \sz_{i} \rangle \langle \sz_{j}\rangle$
where $i$ and $j$ are lattice sites of the two spins.
When quenching from a product state where initially $C_{ij} = 0$ for all $i \neq j$, the time evolution of $C_{ij}$ quantifies the spread of an excitation in time.

The entanglement of two parts of the system is quantified through the von Neumann entropy
$S = - \sum_i \alpha_i^2 \ln \alpha_i^2$,
where $\alpha_i$ are the singular values of the Schmidt decomposition of the chosen bipartition.
$S$ is also a measure of the spread of excitations across the bipartition, as this process entangles the two parts of the system.

\begin{figure}
    \centering
    \includegraphics[width=1.\columnwidth]{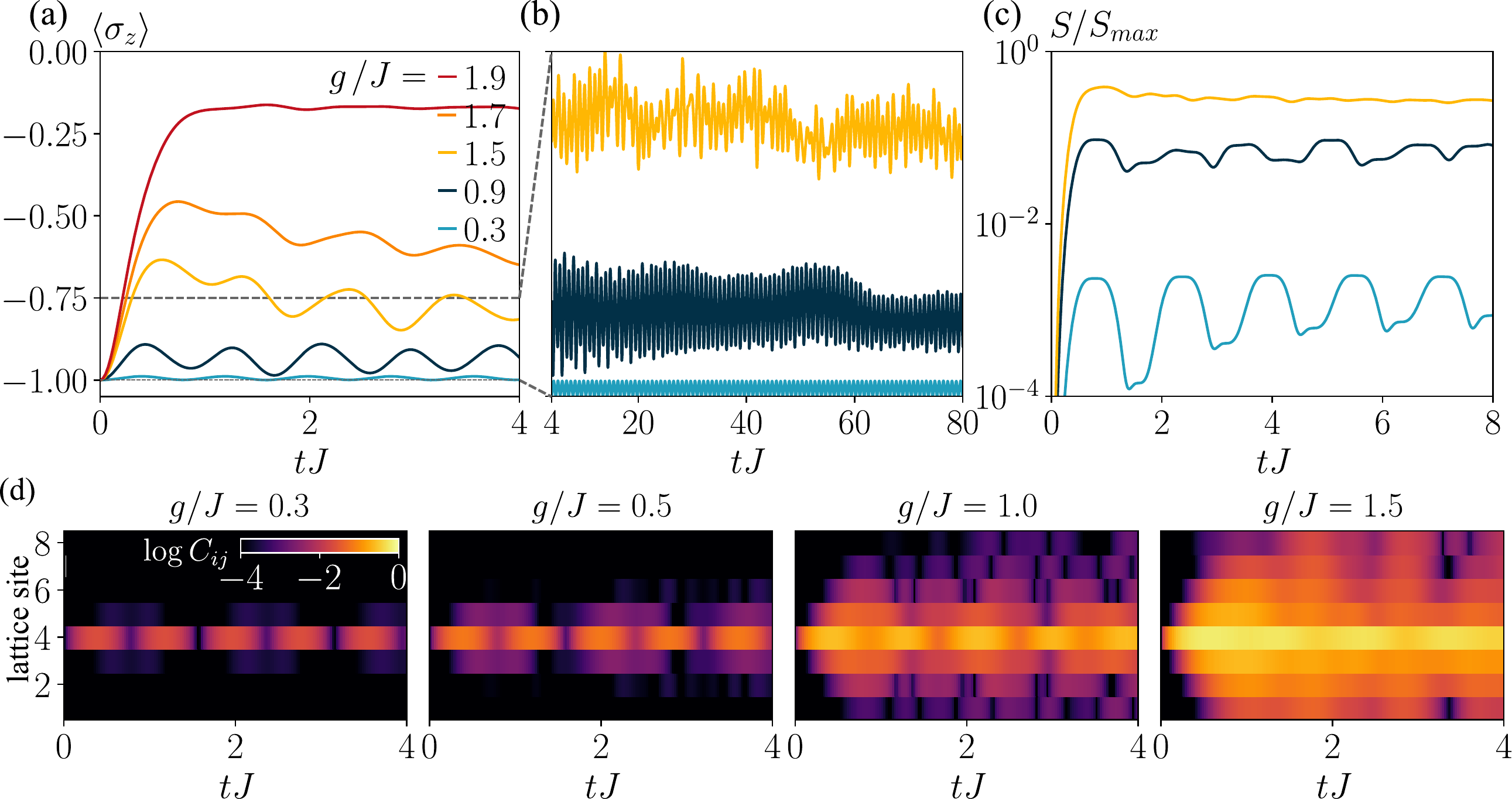}
    \caption{
    Confinement in the 2D quantum Ising model.
    (a) Short time dependence of magnetization $\ev{\sz(t)}$ for a set of $g/J$.
    (b) $\ev{\sz(t)}$ for longer times and small $g/J$. 
    (c) Entanglement entropy $S$ for a bipartition of two neighbouring spins and the rest of the system. $S_\mathrm{max}$ is the maximal entanglement entropy of the bipartition (in this case $S_\mathrm{max} = 2 \ln 2$).
    (d) Horizontal cuts of $C_{ij}$ through the middle of the system.
    }
    \label{fig:nothing}
\end{figure}

\textit{Confinement} ---
Slow thermalization of a many-body system is reflected in persistent oscillations of local observables.
Signatures of confined dynamics are the suppression of the light-cone spread of correlations, and slow and non-monotonous growth of entanglement with time~\cite{Kormos2016}.
We demonstrate these properties after a quench of a completely polarized product state $\ket{\downarrow \dots \downarrow}$ to finite $g$. 
The time dependence of the magnetization $\ev{\sigma_z}$ is presented in Fig.~\ref{fig:nothing}(a,b).
For small $g/J \lesssim 1.5$, we find persistent oscillations with no signs of suppression up to large times. 
However, when increasing $g$ towards $g/J \approx 2$, $\ev{\sigma_z}$ equilibrates on a timescale of $1/J$. This indicates the onset of a regime of quick thermalization, in rough agreement with the critical value of the transverse field for a dynamical phase transition in the same model, $g_c = 2.0J$~\cite{Hashizume2022}.

The time evolution of the entanglement entropy $S$ is shown in Fig.~\ref{fig:nothing}(c). We choose a bipartition between two neighbouring spins and the rest of the system, but all other bipartitions produce similar results.
At small $g/J$, we observe the system periodically returns close to a product state with a frequency of approximately $\frac{\pi}{2} J$, half of the frequency of $\ev{\sigma_z}$. 
The oscillations are washed out as $g$ increases and the density of excitations becomes large enough to interact, leading to thermalization~\cite{Lin2017}.

Fig.~\ref{fig:nothing}(d) shows vertical cuts of $C_{ij}$ with spin $i$ chosen in the fourth row.
The spread of correlations is strongly suppressed for small $g/J$. At $g/J = 0.5$, we find non-monotonous behaviour with the same frequency as the oscillations of entanglement.
The light-cone component becomes more apparent as $g$ is increased, and at $g/J = 1.5$ it is clearly the dominant contribution.
It should be mentioned that the light-cone never vanishes completely, but is only suppressed. Its apparent disappearance is an artefact of the choice of a plotting scale~\cite{Kormos2016, Liu2019}.

\begin{figure}
    \centering
    \includegraphics[width=0.8\columnwidth]{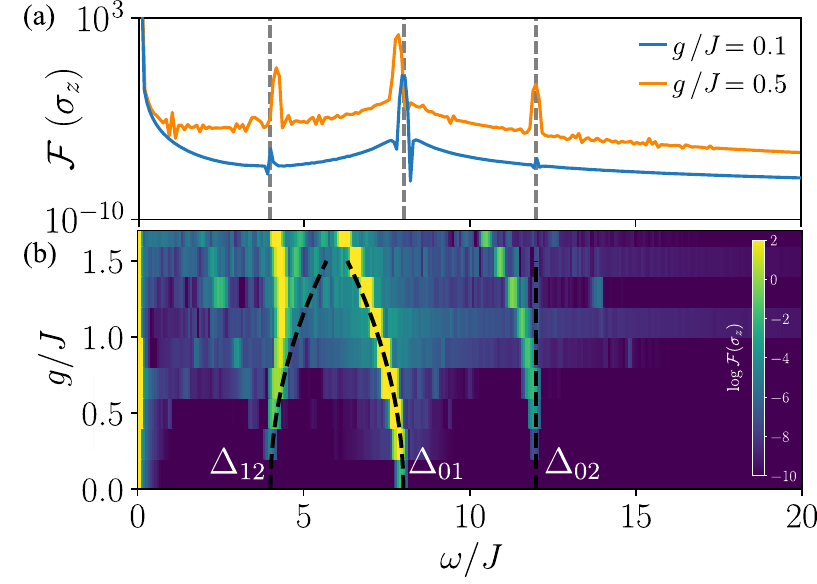}
    \caption{
    Spectrum of $\sz$, given by the Fourier transform $\mathcal{F}(\sz)$.
    (a) Spectrum of magnetization for two cases of $g/J$. Dashed vertical lines correspond to $\omega/J = 4, 8$, and $12$.
    (b) Heatmap of the spectrum for a range of $g/J$. Black dashed lines correspond to the transition energies obtained by perturbation theory. 
    }
    \label{fig:FT}
\end{figure}

\textit{Nature of excitations} ---
The oscillating behaviour of local observables implies that the dynamics is generated by creation and annihilation of local excitations. 
Their nature can be inferred from the spectral density of local observables~\cite{Kormos2016, Lin2017, Liu2019, Tindall2024}. 
In Fig.~\ref{fig:FT} we show the results of the Fourier transform of the magnetization from Fig.~\ref{fig:nothing}(b) between $tJ = 10$ and $tJ = 60$. Altering the upper bound of the interval between $tJ = 30$ and $tJ = 80$ does not change the spectra considerably.
We apply the Hamming window to the data to exclude possible transient events at the edges of the interval~\cite{windowsWiki, Harris1978}.
An example of the obtained spectral density is shown in Fig.~\ref{fig:FT}(a). A heatmap for a broader range is in Fig.~\ref{fig:FT}(b).

The three peaks correspond to excitation energies of the dominant processes.
The most prominent one, at $\omega \approx 8J$, corresponds to the transition between the initial state and a state with one flipped spin, $\psi_1$.
The peak at $12 J$ corresponds to a transition between $\psi_0$ and $\psi_2$, a state where two neighbouring spins are flipped, total domain wall length of $6$.
Finally, the peak at $4 J$ corresponds to the transition between $\psi_1 $ and $\psi_2$.

The transition energies are captured by perturbative corrections to the product states. 
Up to the second order in $g$ we find:
\begin{equation}
\begin{split}
    E_0 &= -\frac{g^2}{8J} N^2 + \mathcal{O}(g^4), \\
    E_1 &= 8 J -\frac{g^2}{8J} \left( N^2 + 6 \right) + \mathcal{O}(g^4), \\
    E_2 &= 12J -\frac{g^2}{8J}N^2 + \mathcal{O}(g^4).
\end{split}
\label{eq:perturbation_results}
\end{equation}
Note that $N$ is the linear size of our system, consisting of $N^2$ sites in total.

The black dashed lines in Fig.~\ref{fig:FT}(b) indicate the transition energies, $\Delta_{ij} = E_j - E_i$.
They match the positions of the spectral peaks to surprisingly large values of transverse field, up to $g \sim J$.
This agreement is surprisingly good for expressions of second order in $g$, and indicates that the contribution of the higher excited modes is negligible throughout the confining regime.

See the End Matter section for a deeper interpretation of the obtained expressions, and the SM for complete calculations.

\textit{Interfaces} ---
Next, we probe the dynamics of interfaces by investigating quenches of inhomogeneous initial states.
We consider two initial states: \textit{the stripe}, where two domains are separated by a flat interface, and \textit{the square}, where the flipped spins are arranged into a square domain.

The presence of dynamical constraints importantly affects interface dynamics.
Whereas interfaces are typically expected to diffusively melt, in the confining phase they tend to remain stable up to very long times.
This is because classically resonant processes dominate the dynamics for small $g/J$, ie. only spins with two neighbours of each spin species can flip, as the initial and final state have the same energy.
These processes only reshape the domain wall, and thus its total length emerges as a dynamical constraint.
The interface dynamics is well understood in the limit of infinite $J$, where non-resonant processes are explicitly forbidden~\cite{Balducci2022, Balducci2023}.
Slow thermalization comes as a consequence of the subsequent fragmentation of Hilbert space into sectors labeled by the domain wall length, and further into so-called dynamically decoupled Krylov subspaces; two domains cannot merge through local interface reshaping if they are spatially separated~\cite{Yoshinaga2022, Hart2022}.

\begin{figure}
    \centering
    \includegraphics[width=1.\columnwidth]{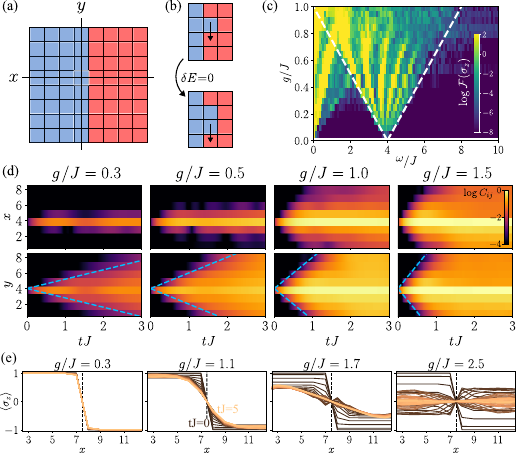}
    \caption{
    Spread of correlations near an interface.
    (a) Sketch of the initial state, with blue representing $\uparrow$ and red $\downarrow$ spins. Black lines indicate the cuts shown in (d).
    (b) Sketch of the resonant process along the interface.
    (c) Heatmap of the spectral density of $\ev{\sz}(t)$ for a spin at the interface for a range of $g/J$.
    White dashed lines correspond to the transition energies of a freely propagating edge mode.
    (d) Horizontal ($x$) and vertical ($y$) cuts of the connected correlations function $C_{ij}$ with respect to the spin at $(4,4)$ (white square in (a)). 
    Blue dashed lines are $\pm 4gt$, showing that the interface mode carries the correlations.  
    (e) Cuts of magnetization along a row in a $16 \times 8$ system, with periodic boundary conditions along the shorter, and open along the longer dimension.
    The system is initialized with two polarized domains and an interface between spins $7$ and $8$ (black dashed line). The cuts are taken at evenly spaced times between $tJ=0$ (darkest brown) and $tJ=5$ (lightest yellow).
    }
    \label{fig:stripe}
\end{figure}

Results for the stripe are presented in Fig.~\ref{fig:stripe}, with a sketch of the initial state shown in Fig.~\ref{fig:stripe}(a).
Recall the periodic boundary conditions, which ensure that the interface is completely flat with no kinks.

The lowest excitation above the stripe is a spin flip at the interface. It creates two additional domain walls and costs $2 \times 2J$. 
Once this excitation is present, flipping the neighbouring interface spin does not come with any additional energy penalty, see Fig.~\ref{fig:stripe}(b).
Such excitations are thus allowed to freely propagate along the boundary, with the expected dispersion $E(k) = 4 J \left( 1 + g \cos(k) \right)$~\cite{Mbeng2020}.

The spectral density of $\ev{\sz}$ for a spin at the interface is shown in Fig.~\ref{fig:stripe}(c). 
The most prominent feature is at $\omega = 4J$ for $g=0$ and fans out into a continuum with increasing $g$, as expected of a freely propagating mode.
White dashed lines correspond to the expected envelopes of the transition energies $\omega = 4J \pm 4g$.  
The continuum is split into discrete peaks as a consequence of the finite size of the lattice.

Vertical and horizontal cuts of the connected correlation functions are shown in Fig.~\ref{fig:stripe}(d).
Because of the interface, the spread of correlations is not the same in the horizontal (x) and vertical (y) directions.
In the $x$ direction (orthogonal to the interface) correlations show signatures of confinement equivalent to Fig.~\ref{fig:nothing}. Interestingly, the correlation spread is symmetric, meaning that the confined excitations spread across the interface into the domain with opposite polarisation in the same way as within the domain. 
On the other hand, the $y$ cuts show very different behaviour. 
The freely propagating interface mode spreads the correlations along the interface with a velocity proportional to $g$ (blue dashed lines in Fig.~\ref{fig:stripe}(d)), even for $g \ll J$.
As $g$ is further increased, the correlations gradually lose their $x$-$y$ asymmetric nature, and at large $g$ spread equally in either direction. 

Due to the scale separation between spin-flips at the interface and in the bulk the crossover from constrained to thermalizing dynamics exhibits an intermediate regime, where the interface broadens while the bulk remains inert. 
While it only faintly appears in the correlations, the regime is clearly visible in the magnetization cuts, Fig.~\ref{fig:stripe}(e). Here we use a $16 \times 8$ system on a cylinder (periodic boundary conditions along the shorter and open along the longer dimension).

The three regimes are discerned by the width of the active area.
Deep in the confining regime (see $g/J=0.3$) only the interface spins are active through the edge mode mechanism.
At large $g$ ($g/J=2.5$), the interface does not present an energetic barrier, and the thermalization occurs indiscriminately at the interface and in the bulk. 
In the intermediate regime ($g/J \sim 1 - 2$), we find that the interface melts and broadens, while the bulk remains inert.
This results in a gradient of magnetization between the two domains at late times.   
A larger set of cuts in the intermediate regime is provided in the SM.

The presence of multiple regimes affects the course of thermalization and the final thermal state. 
Each process has its associated energy scale, and thus a corresponding thermalizing time scale. 
In such cases, the thermalization is expected to occur in stages, as the system relaxes to the Gibbs ensemble corresponding to the subspace of the Hilbert space accessible at a given timescale~\cite{Birnkammer2022}.
This can affect the nature of the final thermal state~\cite{Zadnik2023}.
The influence of the domain wall shape on the thermalization remains an interesting open question.

\begin{figure}
    \centering
    \includegraphics[width=1.\columnwidth]{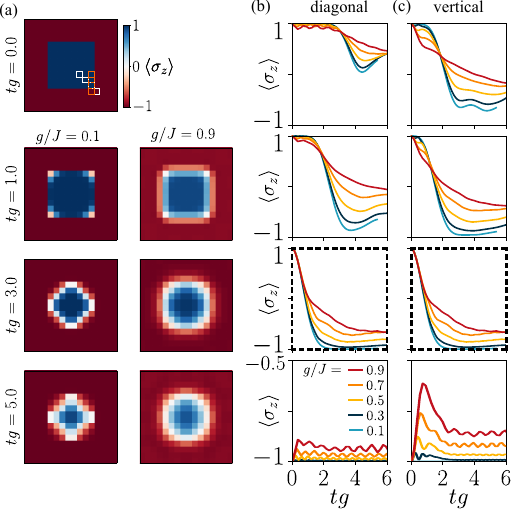}
    \caption{
    Melting of a square.
    (a) Snapshots of magnetization for small and intermediate $g/J$.
    (b,c) Magnetization $\ev{\sz}$ for a set of spins indicated by the (b) white and (c) orange squares in the initial state in (a). The panels with thick dashed frames show the same corner spin.
    }
    \label{fig:square}
\end{figure}

Finally, we consider the melting of an $8 \times 8$ square domain embedded in a $16 \times 16$ lattice.
Fig.~\ref{fig:square}(a) shows the evolution of the magnetization for small and intermediate $g/J$, while Fig.~\ref{fig:square}(b) shows the time dependence of the magnetization for a set of spins close to a corner of the square.
For extended plots of the magnetization as well as the correlation spread, see the SM.

In the $J \rightarrow \infty$ limit and for an infinitely large corner, the melting process can be understood as a series of resonant spin-flip processes~\cite{Balducci2022, Balducci2023}. 
The corner spin has two neighbours of each spin species, so it flips on a timescale set by $1/g$. 
After the it is flipped, its two neighbours resonantly flip as well. This triggers a propagating edge mode in the same way as in Fig.~\ref{fig:stripe}. 
The process results in the square melting into a cross-like shape, as shown in Fig.~\ref{fig:square}(a) for $g/J=0.1$.

A more quantitative view is presented in Figs.~\ref{fig:square}(b,c). 
Due to the dominating edge mode at small, $g/J$ the spins along the interface (Fig.~\ref{fig:square} (c)) melt with a delay which is linear in their distance from the corner.
Probing deeper into the square along the diagonal (Fig.~\ref{fig:square}(b)), we find that spins near the center stay inert up to relatively long times. 

Increasing $g$ causes deviations from this simple picture and a crossover towards a regime where all spins at the interface melt simultaneously.
The amplitude of magnetization oscillations is gradually reduced, and at $g/J = 0.9$ we find a monotonous decrease of $\ev{\sz}$ for all spins of the square simultaneously. 
The melting produces waves of magnetization radiating from sides of the square into the environment (see bottom panel of Fig.~\ref{fig:square}(c)). This is a sign that the dynamics is no longer confined, and the length of the domain wall not conserved. 

\textit{Realization on a quantum simulator} ---
The Ising model is naturally realized on various experimental platforms. 
However, being two-dimensional, the presented setup is most directly applicable to neutral atom arrays and within reach of existing experiments.

More concretely, Ref.~\cite{Scholl2021} demonstrates control over 2D arrays of hundreds of Rydberg atoms realizing a nearest-neighbour antiferromagnetic Ising model with the interaction on the order of $2\,$MHz and coherence time of $20 \mu$s.
When driving the transition between the two spin states at $1\,$Mhz, the timescales shown in Fig.~\ref{fig:square} correspond to experimental times of a few microseconds.
If one chose to relax the nearest neighbour constraint the atoms could be moved much closer together, leading to stronger interactions and thus much faster dynamics. 
See for example Refs.~\cite{Bluvstein2021, Manovitz2024} for experimental and Ref.~\cite{Samajdar2024} for a similar theoretical proposal.

\textit{Outlook} ---
We used tree tensor networks to study constrained dynamics in the ordered phase of the 2D quantum Ising model. 
We showed the signatures of confinement of quasiparticles and extracted the spectrum of excitations from oscillations of the magnetization. The spectrum can be adequately reproduced by perturbation theory up to $g \sim J$. 
We investigated the melting of interfaces between domains, how it depends on confinement and how it changes with increasing transverse field.
We find a crossover between the constrained dynamics of classically resonant processes in the limit of small $g/J$ and diffusive melting at large $g/J$. 

An important implication of the presented work is the fact that these classical simulations are possible up to such long times. 
The underlying reason is slow entanglement growth and thus slow increase of bond dimension required to retain the accuracy of tensor network states. 
The confining regime is thus perfect for benchmarking quantum machines with tensor network calculations~\cite{Tindall2024_IBM}.

Given the demonstrated success of TTNs to accurately simulate the time evolution of dynamically constrained two dimensional systems, we see the potential to tackle various open questions and new avenues of research.

Being very similar in nature to spin systems, the 2D Rydberg arrays in the checkerboard phase also exhibit constrained dynamics~\cite{Manovitz2024}. 
However, the nature of the dynamical constraints, and the properties of basic excitations in such systems are not well understood. 
Investigation of these properties, as well as their interplay with coarsening of domains are well in reach of the presented method, and subject of future work. 

Another promising avenue is the investigation of false vacuum decay in 2D. 
In 1D, the proposed mechanism for the decay is the nucleation of true vacuum bubbles~\cite{Lagnese2021, Pomponio2022, Vodeb_2024}.
Nucleation is driven by the competition between the bulk and interface contributions to the bubble energy. 
In 1D, the area of the bubble remains constant as it grows, while in higher dimensions it increases with the power of $D-1$. 
Exploring nucleation in 2 dimensions should thus provide a qualitatively new, and more universally accurate understanding of the dynamics of the false vacuum.   
A relatively simple starting point would be to investigate how the melting in Fig.~\ref{fig:square} depends on longitudinal field, which makes the square either a false or a true vacuum domain.

Expanding on this idea, it would be interesting to probe the scattering of false vacuum bubbles. 
Such calculations have been performed in one dimension~\cite{Milsted2022}, but multiple scattering channels are expected to arise in two dimensions.
However, clearly discerning scattering from lattice effects requires a significant increase of the system size, possibly beyond the reach of classical simulations.

\acknowledgments
We acknowledge enlightening discussions with Marcello Dalmonte, Roberto Verdel, Simone Notarnicola, Jernej Rudi Finžgar, and Sourav Nandy.

The authors acknowledge funding from 
CN00000013 - Italian Research Center on HPC, Big Data and Quantum Computing (ICSC), 
the H2020 projects EuRyQa, 
QuantERA2021 project T-NiSQ, 
the Quantum Technology Flagship project PASQuanS2, 
the Italian Departments of Excellence grant 2023-2027 Quantum Frontiers, 
the German Federal Ministry of Education and Research (BMBF) the funding program quantum technologies, 
project QRydDemo, 
and the World Class Research Infrastructure - Quantum Computing and Simulation Center (QCSC) of Padova University.

The simulations were performed using the Quantum Green Tea software version 0.3.23 and Quantum Tea Leaves version 0.4.46 \cite{QuantumTea2024}.
The authors acknowledge computational resources of the HPC center Vega at the Institute of Information Science (IZUM) in Maribor, Slovenia and HPC Leonardo by Cineca, Italy.

\textit{End matter on perturbative calculations} ---
In the main text (Eq.\eqref{eq:perturbation_results}) we present perturbative expressions for the energies of the first three excitations above the completely polarized initial state.
Here, we comment on certain aspects of these result.
We 

Firstly, we find that the predicted excitation energies match the calculated spectral excitations up to a suprisingly large $g/J \sim 1$. 
This is an indicator that interpreting the dynamics in the confined regime as isolated excitations above the polarized state is accurate.

Furthermore, the evolution of the spectrum at larger $g/J$ hints at the underlying reasons for the observed quick thermalization beyond $g/J \sim 2$.
$\Delta_{01}$ is the cost of creating a single spin-flip excitation, while $\Delta_{12}$ is the cost of flipping a spin next to an already existing spin flip -- growing a spin cluster by one. Quenching to the point where these two process are energetically equally expensive populates not only the first few but also higher excited states, resulting in the formation of larger clusters. 
At this point, the density of excitations will be large enough for them to overlap and interact, leading to thermalization on a timescale of $1/J$~\cite{Lin2017}.

Note that aside from the extensive correction common to all states, we do not find a second-order contribution to $E_2$. 
An intuitive interpretation is that in second order of $g$ the corrections arising from coupling $\psi_2$ to $\psi_1$, and the ones coming from $\psi_3$ (superposition of connected three-spin clusters, total domain length of 8), originate from the same physical process. This is the addition or removal of a spin from an existing cluster. Both add two domain walls, and being symmetric in this sense, these contributions cancel out.
The processes that couple $\psi_2$ to $\psi_0$ only appear at fourth order, and we expect that accounting for them would reproduce the curvature of the right and left peaks in Fig.~\ref{fig:FT}(b).

Finally, we comment on the differences between the perturbative approach and a similar method of interpreting the spectra -- the effective models. 
Because the dynamics of confined systems is dominated by low energy excitations, it can be accurately described by a simple effective model~\cite{Verdel2020, Liu2019, Tindall2024}.
This is typically done by projecting the Hamiltonian to a subspace of low-energy excitations (equivalents of our $\psi_0$, $\psi_1$ and $\psi_2$). 
However, this approach leads to certain inconsistencies.
Namely, the matrix element between the ground and first excited state is extensive (indeed, in our case we find $\bra{\psi_0}H\ket{\psi_1} = g N$), while others are independent of $N$.
Therefore, an effective model which includes $\psi_0$ predicts a spectral gap (equivalent of our $\Delta_{01})$ that diverges with system size.
To accurately reproduce the excitation energies, the initial vacuum state (equivalent of $\psi_0$) should be excluded from the low energy model, and its energy set exactly to zero.

Perturbative calculations imply that the origin of the problem is in the arbitrary truncation of the subspace of the effective model. This can indeed be read out from intermediate steps of the calculation, see SM.
We find that the extensive corrections come from a state coupling to an extensive number of excitations with one more and one fewer spin flip -- it is always possible to flip a decoupled spin in $\mathcal{O}(N^2)$ sites for an energy penalty of $8J$.
The contributions coming from the state with one more and one less spin flip cancel out in the perturbative calculations, and we only find an overall $g^2 N^2/8J$ correction to all states, so that energy differences remain independent of system size.
This is because the calculation by design accounts for all existing corrections at a given order of $g$.
However, if the Hilbert space is first truncated into an effective model, the coupling between the last retained and the first truncated state is not taken into account. In this case the extensive contributions do not cancel out, and the gaps grow with system size.
While the two approaches produce a similar energy diagram, an advantage of the perturbative approach is that it also provides the correction to the ground state energy. It also allows for direct calculations of the eigenstates.

\newpage
\clearpage

\onecolumngrid

\renewcommand{\theequation}{S\arabic{equation}}
\renewcommand{\figurename}{Supplementary Figure}
\setcounter{equation}{0}
\setcounter{figure}{0}     

\section*{Supplementary Material for: Constrained dynamics and confinement in the two-dimensional quantum Ising model}

\section*{Numerical details}
\label{app:convergence}

The numerical calculations presented in this work are performed with tree tensor networks (TTNs)~\cite{Shi2006, Silvi2019}, implemented in the QTEA library~\cite{QuantumTea2024}. 
In this appendix, we provide qualitative reasoning for the utility of TTNs compared to other tensor networks and technical aspects of the algorithm for time evolution that we apply, and present plots of magnetization and entanglement obtained for a set of maximal bond dimensions.

\subsection*{Tree tensor networks}
In TTNs, the state is represented by a tensor network in the shape of a symmetric binary tree, where the bottom tensors ('leaves' of the tree) carry the physical indices. 
In comparison to the standard matrix product state (MPS) approach, the tree tensor networks are particularly successful at describing states where distant sites are entangled, such as in chains with long-range interactions.
The underlying reason can be understood through a qualitative picture of the way quantum states are encoded in tensor networks.
Qualitatively, entanglement between two correlated physical states is transferred through bonds of the network that connect them. The more entanglement there is, the bigger the bond dimension required to accurately describe the state.
Generally, this also means entangled states are better represented by tensor networks with a bigger number of links.
An extreme example is the projected entangled-pair state (PEPS) ansatz~\cite{PEPS}, where the tensor network mimics the structure of the 2D lattice and each tensor is directly connected to its nearest neighbours.
However, this introduces loops in the network, which leads to technical difficulties. 
Namely, it is not possible to define the isometry center of a network with loops.
Therefore, the algorithms for contracting such networks cannot rely on the underlying symmetries. This leads to poor scaling of calculations of local observables with bond dimension, and limits modern implementations of PEPS to bond dimensions on the order of $\chi \sim 10$. 

The TTN ansatz is in between the MPS and PEPS in the sense that it is the best connected tensor network which by construction has no loops.
For example, two physical sites separated by $m$ lattice sites in real space are only separated by $\sim \ln m$ links in the TTN and $m$ links in MPS. The entanglement between the sites is thus captured more efficiently than in MPS.

\subsection*{Mapping}

An important aspect of TTN simulations in higher dimensions is the mapping of the lattice into a 1D chain, which is then encoded as a TTN.
Throughout this work, we use the Hilbert curve as a mapping function. 
Ref.~\cite{Cataldi2021} shows that this is the optimal choice, in the sense that it maximally preserves locality (the sites which are close in the lattice tend to stay close in the mapped 1D chain). 
We tested various mapping functions, and found that generally the observables converge at smaller bond dimension when using the Hilbert curve.

\subsection*{Time evolution}
The most common way to evolve tensor network states in time is with the time-dependent variatonal principle (TDVP)~\cite{Bauernfeind2020, Haegeman2016}.
The most popular is the two-tensor version of the algorithm (TDVP2), where at each step two neighbouring tensors are contracted, evolved in time, and then split and truncated via SVD. 
This algorithm can grow the bond dimensions between tensors at each time step and is thus a natural choice for simulations starting from weakly entangled states, such as the product states we use.
A much faster version of the same concept is the single-tensor TDVP (TDVP1), where each tensor is optimized without contractions with the neighbours. This approach is much cheaper computationally, but cannot grow the bond dimension of the tensors. This presents a problem if the initial state is a product state with $\chi = 1$.

However, we find that the calculations are much faster if one pads the initial product state with numerical noise on the order of $10^{-16}$ up to bond dimension $\chi$, and performs the evolution with TDVP1, without truncation. 
When comparing TDVP2 with initial bond dimension of 1 and maximal of $\chi=100$ to TDVP1 with constant bond dimension of 100, we observe a speedup in computational time of about $30\times$ for a $8 \times 8$ system.
We also tried hybrid approaches (like performing TDVP2 every $n-$th time step and TDVP1 otherwise, with $n$ around 10).
The advantage of the hybrid approach is that the initial bond dimension is not $\chi$ but rather gradually increases every $n$-th step. This speeds up the initial time steps, but the advantage is quickly lost once the bond dimension reaches a sizeable fraction of the one used for TDVP1. Overall, we still find the TDVP1 to perform faster.

An advantage might be expected for the easiest cases (small $g/J$ limit) where the bond dimension never reaches the maximal allowed value. However, these cases can also be accurately simulated by a TDVP1 simulation with smaller $\chi$, which is anyway much faster.

Another advantage of TDVP1 compared to TDVP2 is that the error of the latter is non-monotonous in the size of the time step. 
In TDVP2, an SVD and a truncation are performed at each time step. This introduces a truncation error, and thus choosing a time step that is too small might actually increase the total error. 
In TDVP1, there is no truncation, and the truncation error is exactly zero.
Consequently, TDVP1 also exactly conserves the energy and norm of the evolved state.

This is an advantage on a technical level as well. Having a single convergence parameter ($\chi$) is more desirable compared to having two ($\chi$ and the minimal truncation $\epsilon$) with non-trivial interplay.

\subsection*{Convergence}
\begin{figure*}
    \centering
    \includegraphics[width=0.8\textwidth]{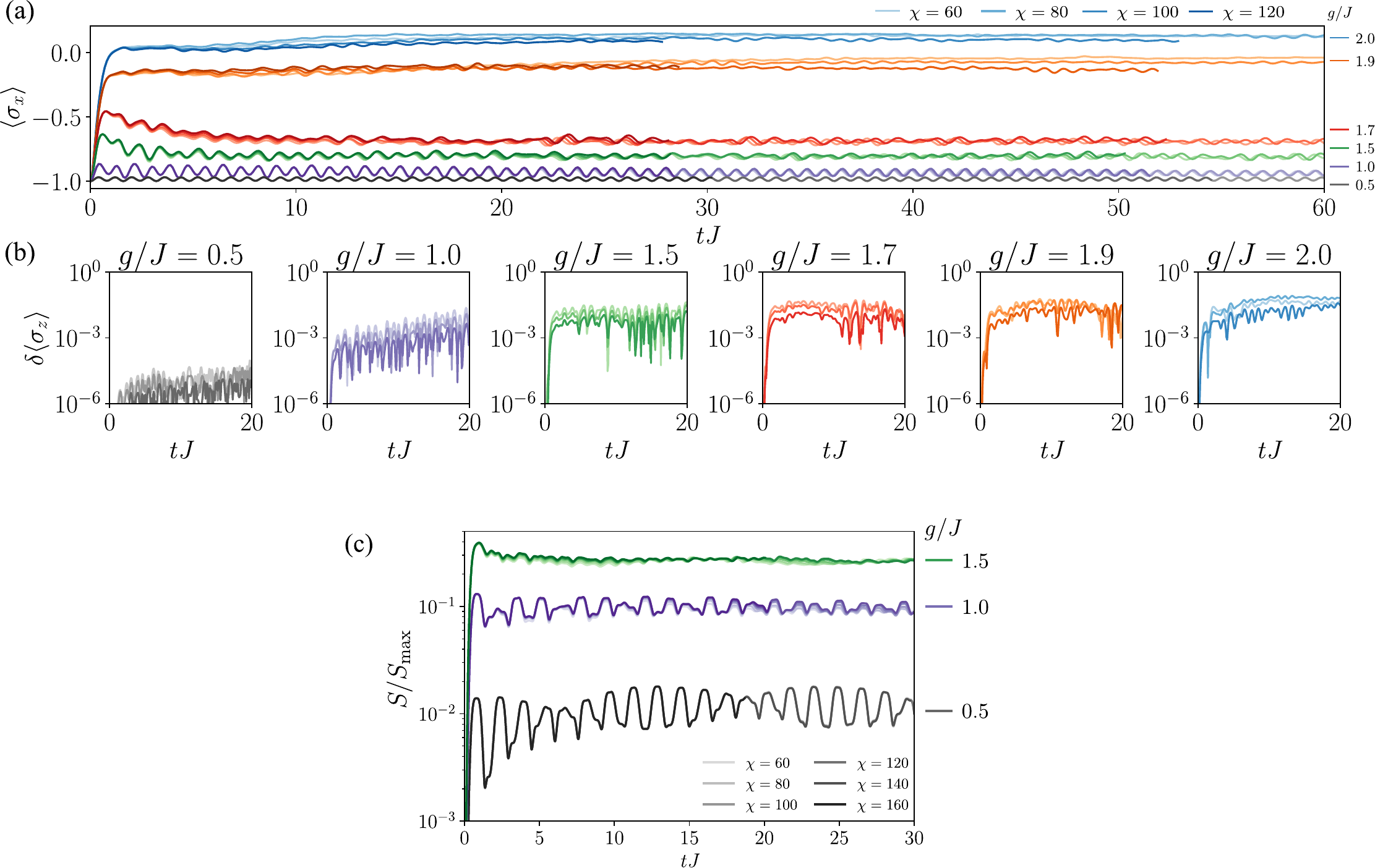}
    \caption{
    Convergence with bond dimension $\chi$.
    (a) Magnetization $\sz$ of a single spin calculated with increasing bond dimension $\chi$ for a set of $g/J$.
    (b) Log scale plots of the deviation of magnetization from the results obtained with $\chi = 160$, $\delta \ev{\sz} = \left\vert \ev{\sz}(\chi) - \ev{\sz}(\chi=160) \right\vert$.
    (c) Entanglement entropy calculated with increasing bond dimension $\chi$ for a set of small $g/J$.
    }
    \label{fig:convergence}
\end{figure*}

In Fig.~\ref{fig:convergence}, we show the convergence of (a) magnetization and (c) entanglement with increasing maximal bond dimension $\chi$. 
The deviation of the magnetization at given $\chi$ with respect to the results obtained with $\chi = 160$ are shown in Fig.~\ref{fig:convergence}(b).

In calculations with tensor networks it is typically observed that local observables converge first, while global quantities require a much more stringent criteria.
Out of the commonly tested quantities, entanglement is typically the last to converge.
Convergence of magnetization and of entanglement thus present the easiest and the hardest convergence test, respectively. 

Unsuprisingly, we find that the convergence is strongly dependent on the underlying physics. 
Deep in the confined phase at $g \ll J$, results are well converged for $\chi = 60$ or less. At larger $g$, larger $\chi$ is required to obtain well-converged results and the simulations become unreliable much sooner in simulated time. 

We use $\chi = 60$ for simulations of longer times, and find the magnetization results reliable up to $g/J \approx 1.5$.
In simulations where we also want to ensure accurate convergence in entanglement and correlations, we use $\chi = 100$. 

\section*{Perturbative corrections}
\label{app:perturbation}

This appendix contains calculations of the perturbative corrections up to the second order in $g$ for the first two excited states above the completely polarized ground state. 
We treat the interaction part as the unperturbed part
\begin{equation}
    H_0 = -J \sum_{\langle i, j \rangle} \sz_i \sz_j,
\end{equation}
and the transverse field term as the perturbation
\begin{equation}
    V = -g \sum_{i} \sx_{i}, 
\end{equation}
with $i$ running across the $N \times N$ spins of the 2D system.

\subsection*{General remarks}

The first-order perturbative corrections to energy of an unperturbed state $\psi$ are given by:
\begin{equation}
    \delta E^{(1)} = -\bra{\psi} V \ket{\psi}.
\end{equation}
In our case, the unperturbed states are product states of spins polarized in $z$, while $V$ only contains spin flipping terms $\sx$.  
To obtain a non-zero overlap with the initial state, one has to flip at least two spins (or one spin twice), thus the first-order contributions will always be zero. The same argument applies to all odd-order corrections.

In the second order, the correction to a state $\psi$ is
\begin{equation}
\begin{split}
    \delta E^{(2)} &= -\sum_{k \neq \psi} \bra{\psi} V \ket{k} \frac{1}{E_k - E_0} \bra{k} V \ket{\psi} \\
    &= -\sum_{k \neq \psi} \frac{1}{E_k-E_0} \left\vert \bra{\psi} V \ket{k} \right\vert^2,
\end{split}
\end{equation}
where $k$ runs across all unperturbed eigenstates of the system that are not $\psi$, $E_k = \bra{k} H_0 \ket{k}$ being their unperturbed energy and $E_0 =\bra{\psi} H_0 \ket{\psi}$ the unperturbed energy of $\psi$.
In other words, $\delta E^{(2)}$ is the sum of absolute values of all possible off-diagonal matrix elements weighted by the difference of the corresponding state's energy with respect to $\psi$.

In the case of the Ising model all off-diagonal matrix elements are the same, $g$. 
To evaluate $\delta E^{(2)}$, we find all states which are a single spin flip away from $\psi$, group them according to their energy $E_k$, and count the number of such contributions. 

\subsection*{Ground state}

We are looking for the second order correction: 
\begin{equation}
\begin{split}
    \delta E_0^{(2)} &= -\sum_k \bra{\psi_0} V \ket{k} \frac{1}{E_k - E_0} \bra{k} V \ket{\psi_0} \\
    &= -\sum_k \frac{1}{E_k-E_0} \left\vert \bra{\psi_0} V \ket{k} \right\vert^2.
\end{split}
\label{eq:nothing_ground_state}
\end{equation}

The matrix element $\bra{\psi_0} V \ket{k}$ is non-zero only for states with a single flipped spin, $\ket{k} = \sx_{i} \ket{\psi_0}$, with $i$ denoting the coordinates of the flip.
Evidently, there are $N^2$ such states, each with the matrix element $\bra{\psi_0} V \ket{k} = g$ and excitation energy $E_k-E_0 = 4 \times 2J$ coming from the four domain walls.

The sum in Eq.~\eqref{eq:nothing_ground_state} thus gives
\begin{equation}
    \delta E_0^{(2)} = -\sum_{i}^N \frac{g^2}{8J} = - \frac{g^2}{8J}N^2.
    \label{eq:nothing_ground_state_result}
\end{equation}

\subsection*{First excited state}

To find the $g$-dependent correction to the excited states, we repeat the procedure for the state with one flipped spin.
By examining the matrix elements $\bra{\psi_0}  H  \sx_i \ket{\psi_0}$, one notices that the only excited state coupled to $\psi_0$ is the zero-momentum superposition of spin flips
\begin{equation}
    \ket{\psi_1} = \frac{1}{\sqrt{N^2}} \sum_{i} \sx_{i} \ket{\psi_0}.
\end{equation}
It is thus enough to evaluate the perturbative expansion only for this state.
Again the first order correction $\delta E_1^{(1)} = 0$,
while in second order we have
\begin{equation}
    \delta E_1^{(2)} = \sum_k \frac{1}{E_k - E_1} \left\vert \bra{\psi_1} V \ket{k} \right\vert^2,
\end{equation}
with $E_1 = 4 \times 2J$.

In this case, the non-zero contributions come from states with two spin flips:
\begin{equation}
    \ket{k} = \sx_i \sx_j \ket{\psi_0}.
\end{equation}
Depending on the relative positions of $i$ and $j$, there are three distinct classes of $\ket{k}$:
\begin{enumerate}
    \item $i=j$; $\sx_i \sx_i \ket{\psi_0} = \ket{\psi_0}$. There is exactly one such state, with intermediate energy $E_k-E_1 = -4 \times 2J$.
    \item $i,j$ neighbours, denoted $\ket{\psi_2'}_{ij} = \sx_i \sx_j \ket{\psi_0}$. There are $\frac{1}{2}4N^2$ such states (4 for each lattice site, but $\frac{1}{2}$ for double counting). Flipping neighbouring spins only creates 6 domain walls, thus the intermediate energy of this type of process is $2 \times 2J$.
    \item $i,j$ neither neighbours nor equal, denoted $\ket{\psi_2''}_{ij}$. There are $\frac{1}{2} N^2 (N^2-5)$ such states, $\frac{1}{2}$ again for double counting. The intermediate energy is $4 \times 2J$.
\end{enumerate}
Now, we compute the matrix elements for the three classes.

For $\psi_0$, we have
\begin{equation}
\begin{split}
        \bra{\psi_0} V \ket{\psi_1} &= \bra{\psi_0} -g \sum_i \sx_i \frac{1}{\sqrt{N^2}}\sum_j \sx_j \ket{\psi_0} \\
    &= -\frac{g}{\sqrt{N^2}} \sum_{ij} \bra{\psi_0}  \sx_i \sx_j \ket{\psi_0} \\
    &= -\frac{g}{\sqrt{N^2}} \sum_{ij} \delta_{ij} \\
    &= -\frac{g}{\sqrt{N^2}} N^2,
\end{split}
\end{equation}
and the total contribution to $\delta E_1^{(2)}$ is
\begin{equation}
    \frac{1}{E_0 - E_1} \left\vert \bra{\psi_0} V \ket{\psi_1} \right\vert^2 = -\frac{1}{8J} \times g^2N^2.
\end{equation}
For the basis states with neighbouring flipped spins $k$ and $l$, $\ket{\psi_2'}_{kl}$, we find
\begin{equation}
  \begin{split}
    \bra{\psi_2'}_{kl} V \ket{\psi_1} &= -\frac{g}{\sqrt{N^2}} \bra{\psi_2'}_{kl} \sum_{ij} \sx_i \sx_j \ket{\psi_0} \\
    &= -\frac{g}{\sqrt{N^2}} \times 2,
\end{split}  
\end{equation}
as the sum over $i$ and $j$ gives two ways to match the two flipped spins in $\psi_2'$.
The total contribution of all such states to the energy shift is thus
\begin{equation}
    2N^2 \frac{1}{E_2' - E_1} \left\vert \bra{\psi_2'}_{kl} V \ket{\psi_1} \right\vert^2 = 2N^2 \times \frac{4g^2}{N^2} \times \frac{1}{4J} = \frac{2g^2}{J}.
\end{equation}
Finally, for $\psi_2''$ the matrix element is the same as above:
\begin{equation}
    \bra{\psi_2''}_{kl} V \ket{\psi_1} = -\frac{g}{\sqrt{N^2}} \times 2,
\end{equation}
however, the intermediate energy is $8J$, and the number of states is bigger by a factor of $N^2-5$, producing another extensive contribution to the energy shift:
\begin{equation}
    \frac{1}{2} N^2 (N^2-5) \times \frac{4g^2}{N^2} \times \frac{1}{8J} = 2 \frac{g^2}{8J} \left( N^2 - 5 \right).
\end{equation}
The complete second-order energy correction is the sum of these terms:
\begin{equation}
\begin{split}    
    \delta E_1^{(2)} &= - \left( -\frac{g^2}{8J}N^2 + \frac{2g^2}{J} + \frac{g^2}{8J}2\left(N^2-5\right) \right) \\
    & = -\frac{g^2}{8J} \left( N^2 + 6 \right).
\end{split}
\end{equation}
Importantly, the extensive ($\propto N^2$) contribution is the same as for the ground state, see Eq.~\eqref{eq:nothing_ground_state_result}, resulting in a $N$-independent gap.

\subsection*{Second excited state}

The second excited state is of type $\psi_2'$, with two neighbouring spins flipped. 
Again, it is enough to consider the zero-momentum superposition
\begin{equation}
    \ket{\psi_2'} = \frac{1}{2N^2} \sum_{ij} \sx_i \sx_j \ket{\psi_0},
\end{equation}
with $i$, $j$ neighbours.

The relevant classes of states coupled to $\ket{\psi_2'}$ have three spin flips over the ground state, $\sx_i \sx_j \sx_k \ket{\psi_0}$. They are:
\begin{enumerate}
    \item two of $i$, $j$, $k$ equal, producing a state with a single flipped spin, $\ket{\psi_1}$. There is $N^2$ such states, and the intermediate energy is $-2 \times 2J$.
    \item $i$, $j$, and $k$ forming a connected three spin cluster, the state denoted $\ket{\psi_3'}=\sx_i \sx_j \sx_k \ket{\psi_0}$. There are $6  N^2$ such clusters on a $N \times N$ lattice. The intermediate energy comes from adding two domain walls, and is thus $2 \times 2J$.
    \item $k$ disconnected from $i$ and $j$, denoted $\ket{\psi_3''}$. There are $2N^2 (N^2-8)$ such states -- for each $i$, $j$ cluster the disconnected spin can be flipped on the remaining $N^2-8$ sites.    
\end{enumerate}

For the first case, we find the matrix element
\begin{equation}
    \bra{\psi_1} V \ket{\psi_2'} = \bra{\psi_0} \sx_i \sum_j g \sx_j \frac{1}{\sqrt{2N^2}}\sum_{kl} \sx_k \sx_l \ket{\psi_0} = g\frac{4}{\sqrt{2N^2}}.
\end{equation}
In words, there are four ways to match any one flipped spin by starting from a superposition of all possible two-spin clusters and flipping one of them.
The $\psi_1$ contribution to the correction is thus 
\begin{equation}
    N^2 \times \frac{-1}{4J} \times \frac{16g^2}{2 N^2}.
\end{equation}

In the second case, the matrix element is
\begin{equation}
\begin{split}    
    \bra{\psi_3'} V \ket{\psi_2'} = g\frac{2}{\sqrt{2N^2}}.
\end{split}
\end{equation}
Each connected three-spin cluster can be constructed by flipping a neighbouring spin of two distinct two-spin clusters.  
The total contribution is
\begin{equation}
    6N^2 \times \frac{1}{4J} \times \frac{4g^2}{2 N^2}.
\end{equation}

Finally, the case of a decoupled flipped spin is similar to the third case of the $\psi_1$ corrections, but here with a different geometrical factor. 
The matrix element is 
\begin{equation}
    \bra{\psi_3''} V \ket{\psi_2'} = g\frac{1}{2N^2},
\end{equation}
and thus the total contribution
\begin{equation}
    N^2 \left( N^2 - 8 \right) \times \frac{1}{4J} \times \frac{g^2}{2N^2}.
\end{equation}

When summing all contributions all $N$-independent terms cancel out, and we obtain the second order correction to the energy of $\psi_2'$:
\begin{equation}
    \delta E_2^{(2)} = -g\frac{g^2}{8J} N^2.
\end{equation}
See the main text around main text Fig.~2 for a qualitative interpretation of the obtained result.  

\FloatBarrier
\section*{Additional results for the interface}
In this section we provide additional results of the time evolution of magnetization in the case of an interface.
These are obtained in a $16 \times 8$ system on a cylinder (periodic boundary conditions along the shorter and open boundary along the longer dimension).

As described in the main text, we find three dynamical regimesl, and smooth transitions between them.
For small $g/J$, the system is constrained. The spins at the interface oscillate via the edge mode.
At intermediate $g/J \sim 1 -- 1.7$, the dominant process is the interface broadening, while the rest of the system remains static. 
At large $g/J$ the dynamical constraint does not play a role. All spins oscillate simultaneously with a frequency of approximately $J$, and the system thermalizes quickly.

\begin{figure*}
    \centering
    \includegraphics[width=1.\textwidth]{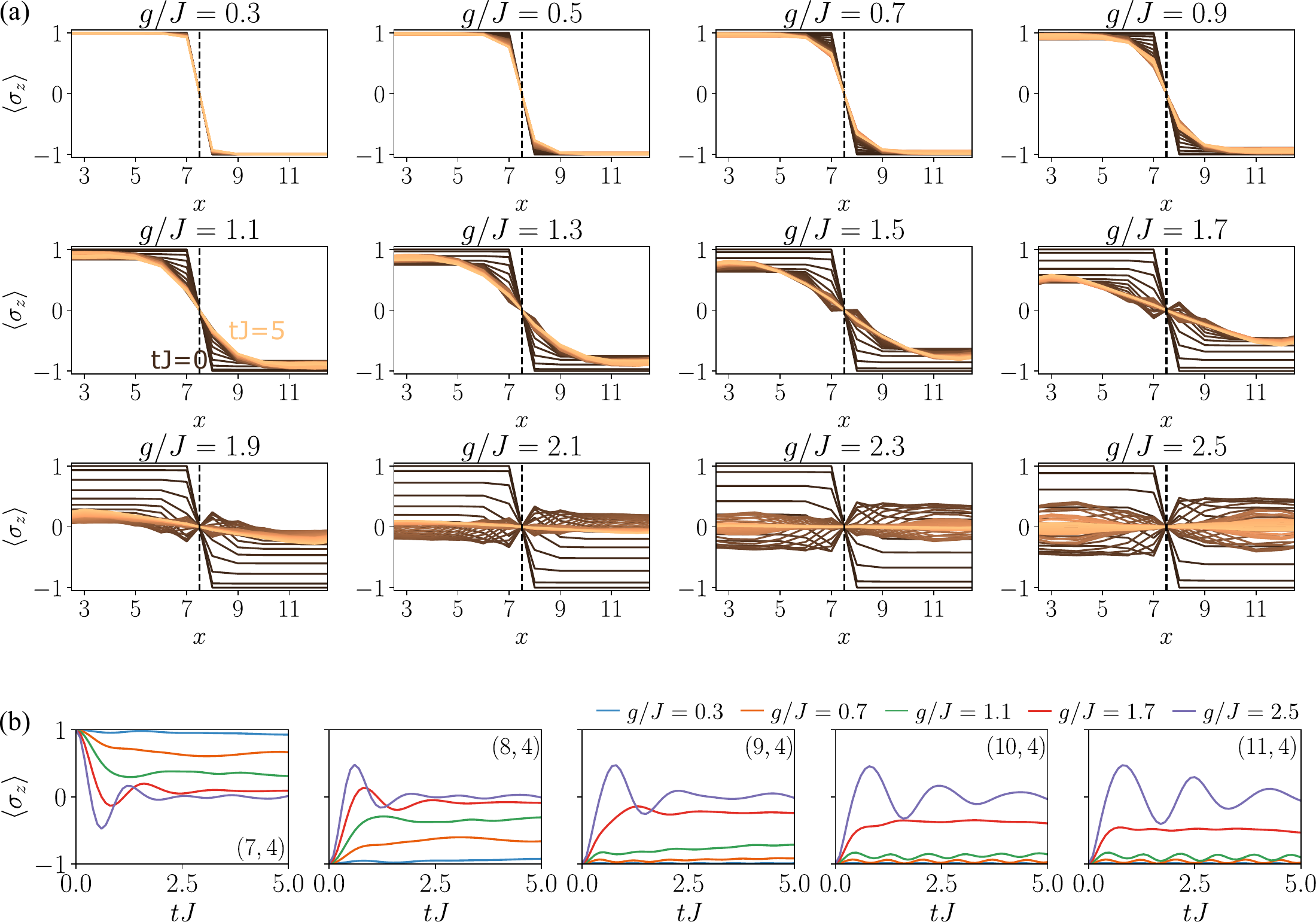}
    \caption{
    Time evolution of the magnetization $\sigma_z$ for a stripe initial state; $16 \times 8$ system, initialized with two domains. The interface is between the $7$th and $8$th spin in the longer dimension. 
    (a) Cuts of magnetization for an extended set of $g/J$.
    (b) Magnetization for five spins, one in the $\uparrow$ polarized domain, and four in the $\downarrow$ polarized domain. The spin coordinates are denoted in the brackets in each panel. 
    }
    \label{fig:sm:square_magnetization}
\end{figure*}

\section*{Additional results for the square}

In this section, we provide additional results for the dynamics of the square.
Fig.~\ref{fig:sm:square_magnetization} shows a grid of magnetization for the lower right quadrant of the square (the panels with blue frames) and two surrounding rows and columns (red frames). Time is rescaled by $g$, as in main text Fig.~5.

As discussed in the main text, we observe a crossover between two limiting behaviours; at small $g/J$ the dynamics consists of a sequence of resonant spin flips which originate from the corner, while with increasing $g/J$ the spins at the interface melt simultaneously.
Spins within the square remain inert until the wave of melting reaches them. 
Interestingly, because the resonant process is a complete spin flip which propagates ballistically, the magnetization at intermediate times ($\sim 6 tg$) generally reaches lower values for small $g/J$, and the square seems to start melting more quickly. 
However, while the small-$g$ dynamics leads to multiple oscillations of magnetization, the large $g/J$ curves monotonously decay to their thermal value.
Longer time simulations are required to accurately determine the nature of the thermalized state.

The diffusive dynamics at large $g/J$ spreads outside of the initial square, while it is confined to it at small $g/J$. 
This is most apparent in the set of spins that neighbour the square, where we find a large initial jump of magnetization (note that the red-framed panels have a $y$-scale from $-1$ to $-0.5$). 
Its magnitude increases approximately linearly with $g/J$. 
For smaller $g/J$ the environment is effectively unperturbed at the scale shown here.

\begin{figure*}
    \centering
    \includegraphics[width=1.\textwidth]{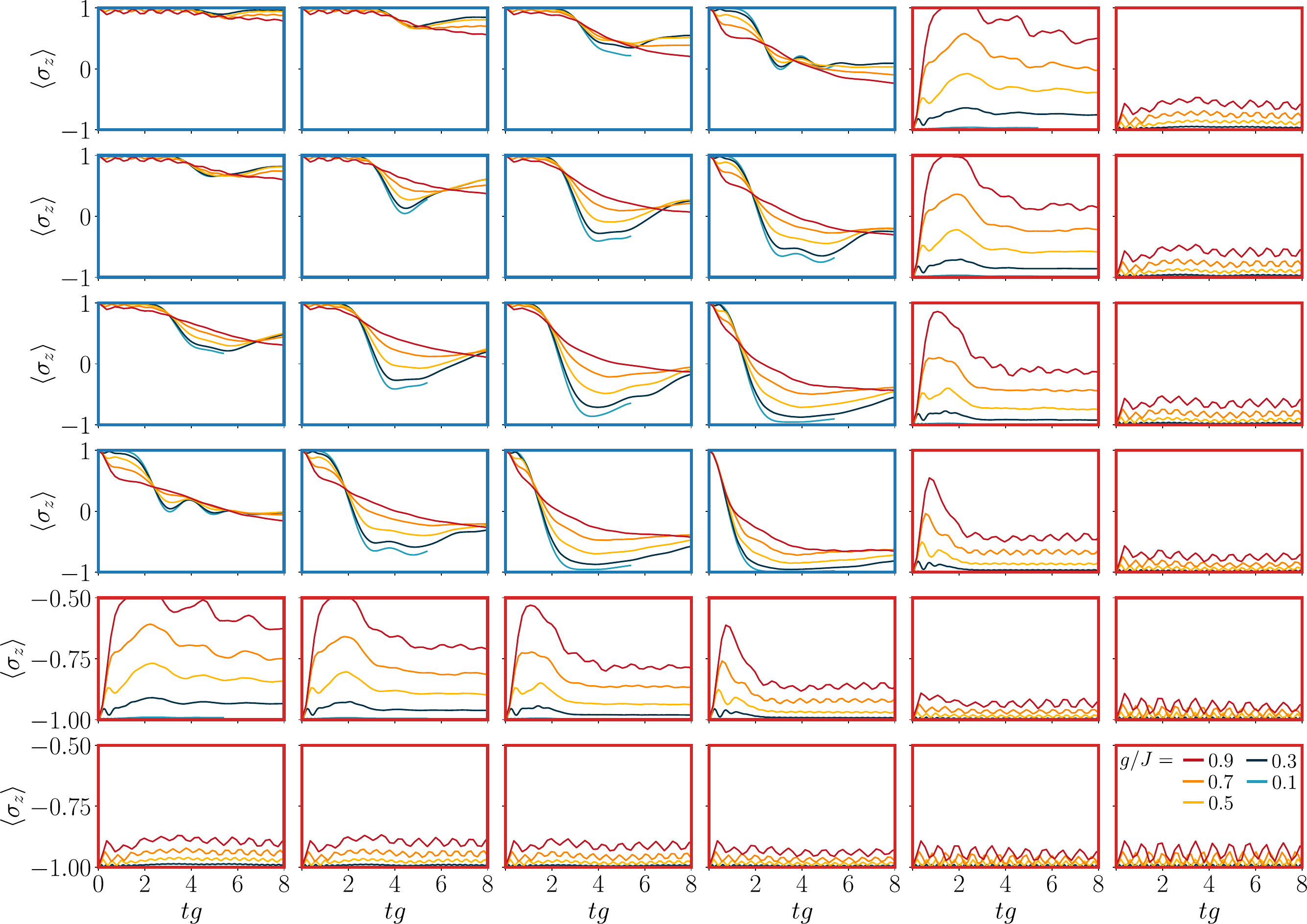}
    \caption{
    Time evolution of magnetization for the lower right quarter of the square (blue frames) and two rows of surrounding spins (red frames).
    The y-range of the blue panels is from $-1$ to $1$, while for red panels it is $-1$ to $-0.5$.
    }
    \label{fig:sm:square_magnetization}
\end{figure*}

\begin{figure*}
    \centering
    \includegraphics[width=1.\textwidth]{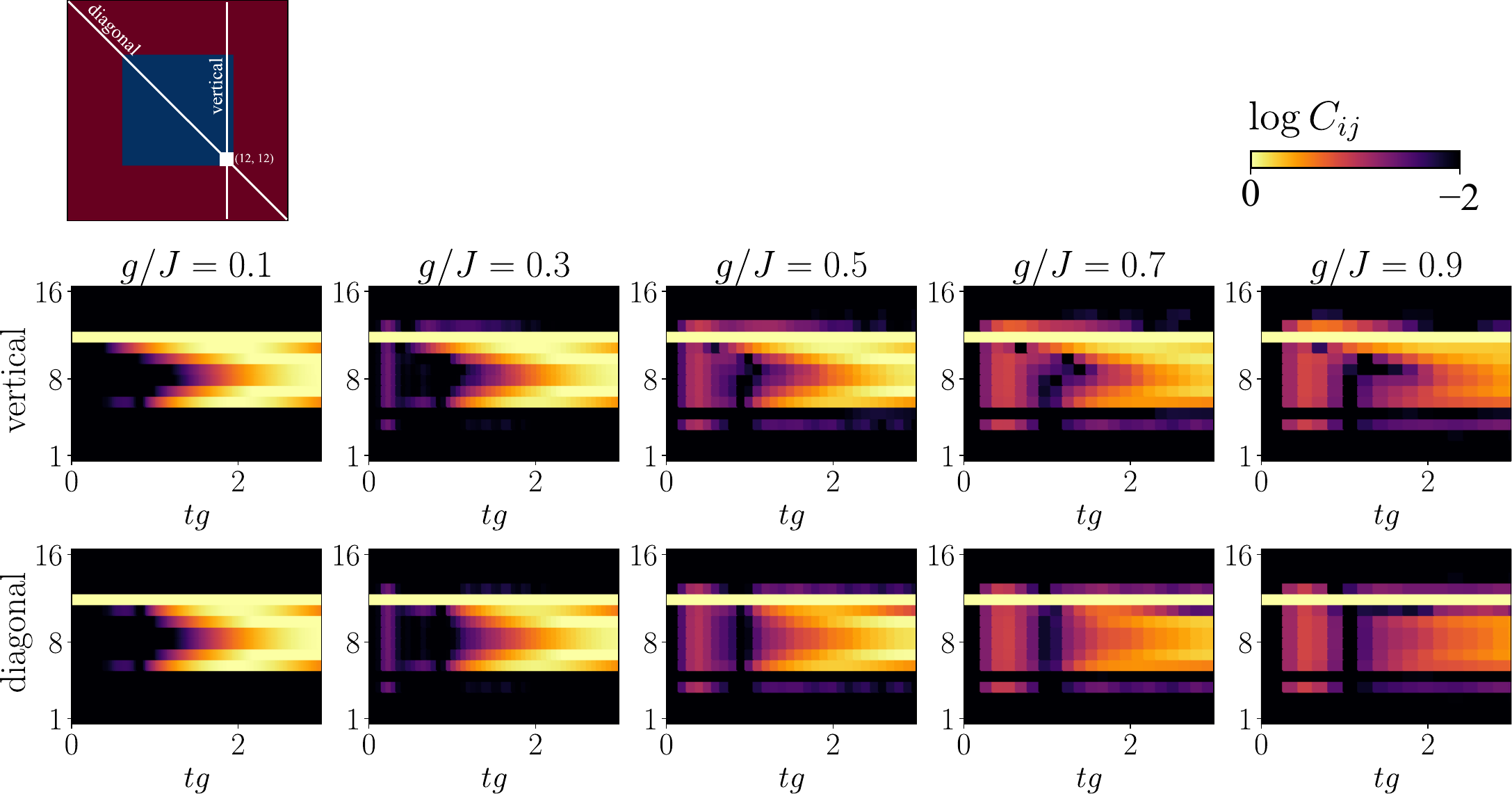}
    \caption{
    Correlations $C_{ij}$ with the $i$-th spin at the corner of the square, at coordinates $(12, 12)$ (counted from the upper left corner as $(1, 1)$). 
    }
    \label{fig:sm:square_correlations}
\end{figure*}

In Fig.~\ref{fig:sm:square_correlations}, we show the correlations $C_{ij}$ where $i$ is the spin in the corner of the square, and $j$ are taken along a vertical (top) or a diagonal (bottom) cut along the system.
For $g/J = 0.1$ we find the expected linear correlation spread along the side of the square, while it is somewhat slower along the diagonal.
Interestingly, because the edge mode propagates much faster, the correlations with the spin in the opposite corner of the square begin growing approximately at the same time as the ones with the neighbouring spin along the diagonal. 

As $g/J$ is increased, we find the expected light-cone spread, and the spread of correlations outside of the initial spread.
An interesting feature that appears at larger $g/J$ are areas where $C_{ij}$ decreases (at $tg \sim 1$). This is probably due to interference of signals propagating along different paths from the corner spin. 

\bibliography{bibliography}

\end{document}